# Design of a monolithic low-threshold narrow-linewidth cw mid-IR silicon Raman laser

Behsan Behzadi, Ravinder K. Jain, Mani Hossein-Zadeh

*Center for High Technology Materials and Optical Science and Engineering program, Department of Physics and Electrical Engineering, University of New Mexico, NM 87131, USA*
*bbehzadi@unm.edu*

**We present the design and anticipated performance of an optically pumped monolithic all-silicon mid-IR (MIR) Raman laser. For the proposed 3.15 µm cw laser design pumped at a 2.71 µm pump wavelength, we estimate a threshold pump power of < 10 mW and an estimated output laser linewidth of <50 MHz. This design may pave the road toward demonstration of the first mid-IR silicon Raman laser.**

In recent years there has been a growing interest in silicon based narrow-linewidth (NLW) mid-IR (MIR) sources, due to their potential as integrated light sources for applications ranging from integrated biological and environmental sensors to free-space communications and LIDAR [1-3]. In the near-IR (NIR) telecom wavelength band (1.2 - 1.7 µm), the absence of a silicon-based laser has been mainly addressed by hybrid integration of silicon devices with those based on III-V semiconductors. In addition, integrated NIR silicon-based Raman lasers have also been developed and studied -- as alternative sources to semiconductor lasers [4] -- in part because of their wavelength flexibility and superior spectral purity characteristics. In particular, several near-IR integrated silicon Raman lasers -- based on low-Q Fabry-Perot cavities [5,6], microring cavities [7] and photonic crystal cavities [8] -- have been demonstrated. In the MIR range, narrow linewidth (NLW) continuous wave (CW) semiconductor lasers are much more expensive and their hybrid integration with silicon has not been reported yet. As such, given the growing interest in MIR silicon photonic circuits for sensing applications, the design, analysis and eventual demonstration of on-chip NLW MIR silicon Raman lasers is very important.

It is well known that silicon is an excellent optical waveguide material at MIR wavelengths > 2.2 µm, since two-photon absorption and three-photon absorption effects are negligible at these wavelengths [9]. Although there have been several theoretical studies and many attempts towards demonstration of MIR Raman lasing in silicon waveguides [10, 11], no experimental demonstrations have yet been reported, mostly due to the absence of low-loss MIR silicon waveguides or the lack of monolithic high-Q cavities (required in conventional designs) and -- to a smaller extent – the lack of MIR pump sources with sufficient power and beam quality to deliver the power levels needed in the silicon waveguides for efficient operation of such on-chip Si Raman lasers. As such, there is still a strong need for practical integrated MIR Raman laser designs that will facilitate operation at low-threshold powers.

In this paper, we propose a novel monolithic design for a MIR silicon Raman laser based on strong narrow linewidth feedback provided by a π-phase shifted waveguide Bragg grating (pps-WBG). The laser is designed for CW pumping in TE polarization at 2.71 µm and generation of Raman laser in TE polarization at 3.15 µm. While the 2.71 µm pump wavelength has been selected due to availability of high power Er:ZBLAN fiber lasers at this wavelength, the same design strategy can be implemented for generating Raman lasers at longer wavelengths using longer pump wavelengths. Using achievable values of waveguide loss (~1 dB /cm), our proposed design should yield a single frequency narrow linewidth (<50 MHz) Raman laser with a threshold power as low as 200 mW. To further decrease the threshold power we propose resonant pump enhancement by combining the pps-WBG cavity with two broadband waveguide Bragg reflectors. The two cavities can be monolithically integrated on a single silicon-on-insulator (SOI) chip using standard silicon processing techniques and the whole device length will be less than 700 µm. Our calculations show that combination of a pps-WBG with broadband WBG-based resonant pump enhancement should result in threshold pump powers ($P_{th}$) of less than <10 mW (based on single-mode SOI waveguides with 1 dB/cm loss). Given that losses as low as 0.1 dB/cm have been already reported -- for multi-mode SOI waveguides – at wavelengths between 3 and 4.5 μm [12], we believe that an optimized SOI-based Raman laser design (that accounts for multimode operation) may operate at pump threshold powers of ≤ 100 μW.

In the following paragraphs, we first analyze the pps-WBG structure as a stand-alone laser, and then add the WBGs to demonstrate ultimate performance. The core of this silicon laser is an integrated pps-WBG that enables resonant enhancement of the nonlinear interaction in a very small mode-volume (~10 μm$^3$). The base waveguide structure used in our design is an air-clad quasi single-mode silicon-on-insulator (SOI) waveguide with a cross section of 0.9×1 μm ($h$ =1 μm, $w$ = 0.9 μm). These dimensions were selected to minimize the propagation loss due to absorption in the buried oxide and side-wall roughness. This relatively large cross-sectional area still supports only one low loss TE mode (the fundamental TE transverse mode) even in the corrugated regions (discussed below, see Fig. 1) where the width is larger than 0.9 mm. The average effective area ($A_{eff}$) of the fundamental mode is 0.98 μm$^2$ corresponding to a confinement factor of Γ = 0.96. The calculated propagation loss (a) due to material absorption for this waveguide is ~0.018dB/cm and ~0.011dB/cm for pump and laser wavelengths respectively (using silica loss from Ref. 13). However, based on current fabrication technologies, the loss is expected to be dominated by scattering (due to the surface roughness).

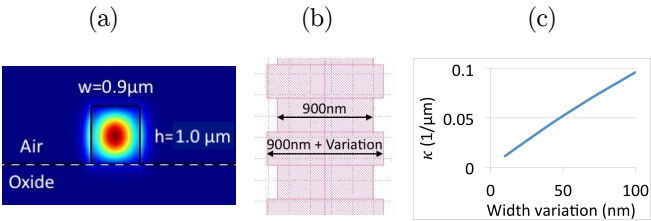

Fig.1: (a) Cross section of the TE optical mode in a "base" SOI waveguide with air cladding. (b) The "top view" of the symmetric grating profile. (c) Calculated coupling coefficient resulting from waveguide width variation.

Fig. 1(b) shows the WBG "grating profile" made by periodic enlargement of the waveguide width from the original width (0.9 μm). The strength of the feedback from this WBG is controlled by the coupling coefficient, $\kappa = \pi\, \delta n_{eff}/\lambda_B$, where $\delta n_{eff}$ is the effective refractive index modulation resulting from variations in the width of the waveguide [14]. Fig. 1(c) shows the coupling coefficient ($\kappa$) plotted against the width variation indicating that κ's as strong as 0.1 μm$^{-1}$ can be achieved by varying the waveguide width by as little as 100 nm. Fig. 2(a) shows the definition of the design parameters for the pps-WBG based silicon Raman laser (only few periods near the p-phase shift region are shown). The operating principles of Raman lasers based on pps-WBGs are well-established and verified in the context of distributed feedback (DFB) Raman fiber lasers [15, 16].

The Bragg wavelength is defined as $\lambda_B = 2n_{eff}\Lambda$, where $\Lambda$ and $n_{eff}$ are the grating period and the effective refractive index of the fundamental waveguide mode respectively. The power build-up factor for the fundamental longitudinal mode of the pps-WBG in the p-phase-shifted region is proportional to $\kappa$ and cavity length ($L_{DFB}$). Using the mathematical framework and methodology for modeling MIR DFB Raman fiber lasers (reported in ref. 16), we have calculated the threshold pump powers ($P_{th,1}$) for operation of a Raman laser at 3150 nm in the fundamental longitudinal mode of the silicon pps-WBG using a 2710 nm pump wavelength. Fig 2(b) shows the calculated contour-plot for $P_{th}$ as a function of $k_{DFB}$ and $L_{DFB}$, assuming the propagation through the grating is limited by absorption loss ($\alpha = 0.011$dB/cm) and that the Raman gain ($g_R$) is $1.0\times10^{-10}$ m/W [17]. Here the nonlinear losses (i.e. two and three photon as well as free carrier absorption) are ignored due to their negligible magnitude in the MIR region at wavelengths above 2.2 μm [9]). As evident from Fig. 2(b), increasing $k_{DFB}$ and $L_{DFB}$ reduces the threshold pump power until it reaches the absorption-limited value (referred to as $P_{th\text{-}min}$) of 3 mW, which is the minimum achievable threshold pump power for this design (assuming $\alpha$=0.011 dB/cm). To evaluate the impact of a non-ideal waveguide loss (> absorption limited), we have calculated $P_{th\text{-}min}$ as a function of a and $A_{eff}$ (see Fig. 2(c)). This figure shows that for waveguide mode with $A_{eff} = 0.98$ μm$^2$, threshold pump power of < 200 mW is achievable even if the propagation loss is as high as 1dB/cm.

Estimating the actual linewidth for a pps-WBG Raman laser is a complicated task (which has not been previously reported). However, its upper limit for the proposed pps-WBG silicon Raman laser can be estimated by the cold cavity linewidth ($P_{pump}<<P_{th}$).

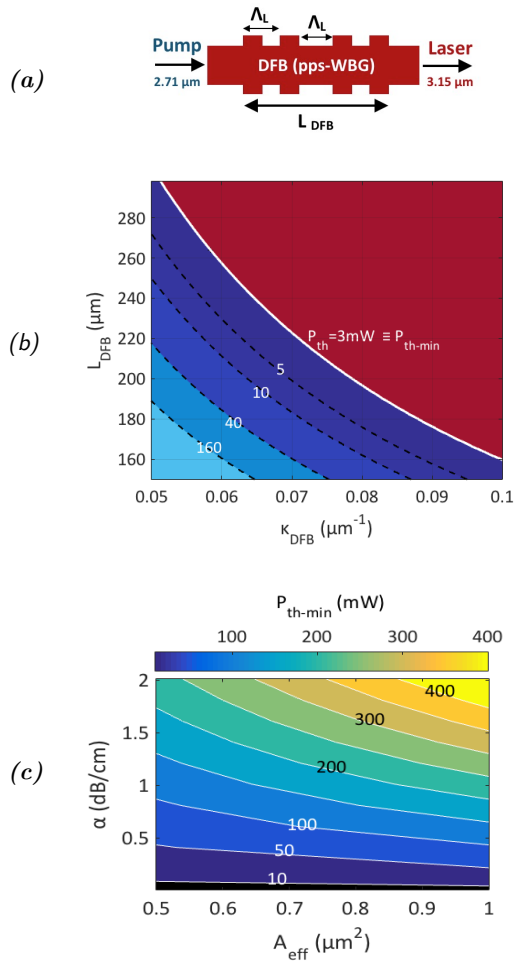

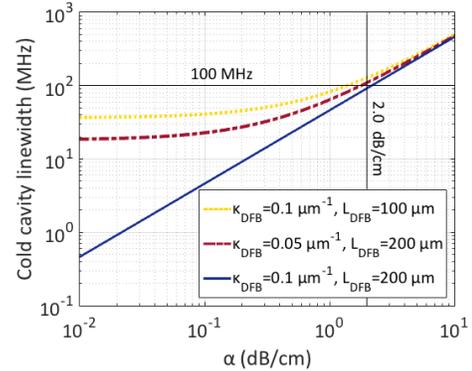

Fig. 2: (a) Schematic diagram of a silicon pps-WBG Raman waveguide laser. (b) Calculated values of $P_{th}$ in a waveguide with a mode area, $A_{eff} = 0.98$ μm$^2$ as a function of $\kappa_{DFB}$ and $L_{DFB}$. Here propagation loss is absorption-limited (0.011 dB/cm) and nonlinear loss is negligible. In this case, the pump wavelength is 2710 nm and the laser (Raman) wavelength is 3150 nm. Variation of $\kappa_{DFB}$ between 0.05 and 0.1 corresponds to a width variation between 46 and 100 nm (see Fig. 1(c)). (c) Minimum threshold pump ($P_{th\text{-}min}$) calculated as a function of as a function of $\alpha$ and $A_{eff}$.

Estimating the actual linewidth for a pps-WBG Raman laser is a complicated task (which has not been previously reported). However, its upper limit for the proposed pps-WBG silicon Raman laser can be estimated by the cold cavity linewidth ($P_{pump} \ll P_{th}$). Fig. 3 shows the cold cavity linewidth plotted as a function of $\alpha$ for three combinations of $\kappa_{DFB}$ and $L_{DFB}$. A linewidth of less than 100 MHz should be achievable for waveguide losses of up to 2.0 dB/cm, which is adequate for most Doppler-broadened atmospheric pressure gas sensing applications where the width of the absorption lines are usually in GHz range. The blue line in Fig. 3 is the absorption limited value of the cold cavity linewidth which can be achieved for 0.01-10 dB/cm loss range with moderate $\kappa_{DFB}$ and $L_{DFB}$ values of 0.1 μm$^{-1}$ and 200 μm respectively.

Fig. 3: Estimated upper limit (cold-cavity) for the linewidth of the pps-WBG Raman silicon laser.

Note that although the pps-WBG can also support higher order longitudinal modes (known as side modes), the frequency of these modes are sufficiently far from the fundamental mode (designed to match with the peak Raman gain) to experience enough gain to grow due to the relatively narrow Raman gain bandwidth of silicon which is about 3-4 nm for a 2.71 μm pump wavelength [18]. Therefore, for a moderate range of pump powers above threshold (up to $P \sim 10 \times P_{th,1}$), the higher order longitudinal modes of the pps-WBG will not experience enough gain to reach the threshold powers $P_{th,2}$ needed for the first higher order mode [15, 16].

To further reduce $P_{th}$, the threshold pump power needed to operate the pps-WBG (designed to sustain the laser wavelength, $\lambda_{laser}$) laser, we propose to resonantly enhance the pump power using two identical broadband WBGs -- designed at a wavelength and bandwidth that matches the pump wavelength ($\lambda_{pump}$) and bandwidth of the pump radiation. As shown in Fig. 4(a), the symmetric WBGs form a Fabry-Perot cavity for $\lambda_{pump}$ cavity to provide maximum field enhancement around the p-phase-shifted-region of the pps-WBG, thereby increasing the efficiency of pumping of the Raman DFB laser. The Bragg wavelength for the DFB ($\Lambda_L$) and pump WBGs ($\Lambda_p$) are equal to $\lambda_{laser}$ and $\lambda_{pump}$ respectively. As such the WBGs (designed based on the same base waveguide structure) provide high reflectivity at the pump wavelength ($\lambda_{pump} = 2.71$ μm) and high transmittance at the Raman laser wavelength ($\lambda_{laser} = 3.15$ μm). Fig 4(b) shows the transmission spectrum of the entire structure (for $\kappa_P = \kappa_{DFB} = 0.05$ μm$^{-1}$, $L_{DFB} = 200$ μm, and $L_p = 100$ μm. $\kappa_p$ is the coupling coefficient for the pump WBGs), showing that the crosstalk between the two cavities is negligible. We have estimated $P_{th,-min}$ for the complete device (Fig. 4(a)) as a function of $L_p$ (by selecting the optimal values for $L_{DFB}$ and $k_{DFB}$). Fig. 4(c) shows the contour-plot of $P_{th-min}$ as a function of the $L_p$ for several values of waveguide loss (assuming equal propagation losses for pump and laser wavelength). Here $\kappa_P = 0.05$ μm$^{-1}$, $\kappa_{DFB} = 0.07$ μm$^{-1}$, and $L_{DFB} = 200$ μm (almost equal to the distance between two pump gratings). The optimal values of $L_p$ that result in the minimum threshold pump powers are depicted in Fig 4(c). These optimal lengths correspond to the critical coupling condition for the pump cavity. For waveguide losses of up to 1dB/cm, the minimum achievable threshold pump power can be as low as 10 mW for $L_p \sim 120$ μm. Moreover, our calculation shows that a $P_{th-min}$ of below 100 mW is simply achievable even using waveguides with higher propagation loss (up to $\alpha_{P,L} = 2$ dB/cm).

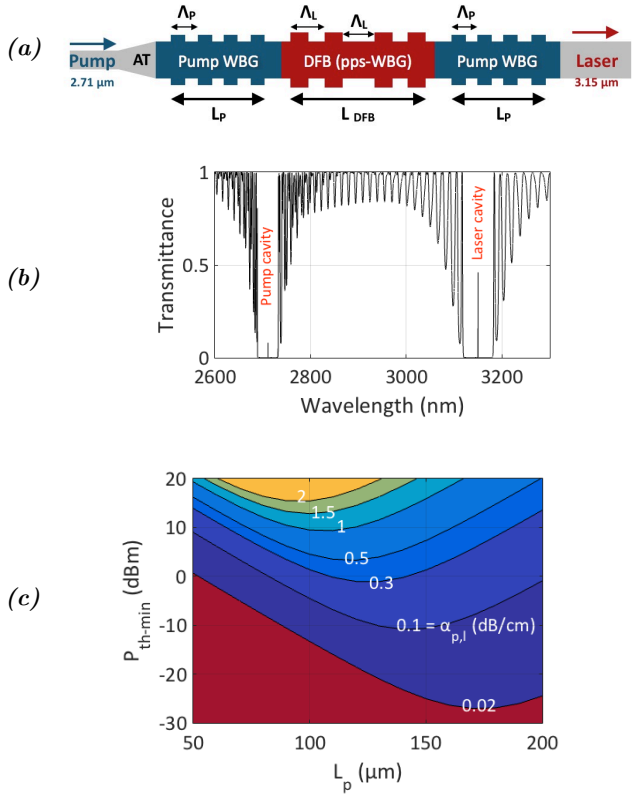

Fig. 4: (a) Schematic structure of the proposed all-silicon Raman laser based on pps-WBG employing resonant pump enhancement. (b) Transmission spectrum through the structure shown in (a) for $\kappa_P = \kappa_{DFB} = 0.05$ mm$^{-1}$, $L_{DFB} = 200$ μm, and $L_p = 100$ μm. (c) $P_{th-min}$ for Raman lasing in the structure shown in (a) as a function of $L_p$ and several values of $\alpha_{l,p}$, (propagation loss for the pump and laser wavelengths are assumed to be equal). Here $\kappa_p = 0.05$ μm$^{-1}$, $k_{DFB} = 0.07$ μm$^{-1}$ and $L_{DFB} = 200$ μm.

In fact, single-mode MIR SOI waveguides with losses below 1.0 dB/cm have already been demonstrated [1,19]. More recently, multimode MIR SOI waveguides with ultralow losses -- of below 0.1 dB/cm -- have been demonstrated [12]. Using a multimode waveguide for fabrication of a DFB Raman silicon laser requires careful design to efficiently deliver the pump power to fundamental TE mode of the waveguide while taking "parasitic effects" in the reflection spectrum of the waveguide Bragg gratings into consideration. For such a design, the pump can be delivered to the fundamental mode of the waveguide using an adiabatic taper [20] to maintain maximum overlap with Stokes (laser) modes and the pump modes. Moreover, the parasitic spectral effect will not degrade the reflectivity of Bragg gratings for the fundamental mode of the waveguide [21], while the relatively large modal dispersion of the waveguide – and the narrow Raman gain bandwidth of silicon naturally results in insufficient gain for higher order modes

Note that as opposed to previous designs (where pump and laser resonance were spectrally coupled [6,7,10]), this structure allows for tuning each separately to compensate the possible fabrication variability. In the above design the pump and laser wavelengths can be changed independently by local thermal tuning of the WBGs or the pps-WBG [22]. Considering the fabrication variability and narrow spectral bandwidth of both pump and laser modes (see Fig.4 (b)) this is an important feature that may also enable a wavelength tunable silicon Raman laser (by tuning the pps-WBG resonance within the Raman gain spectrum).

In summary we presented a CMOS compatible all-silicon mind-infrared Raman laser that can be implemented on SOI substrates using standard fabrication methods. The laser is compact and its total length does not exceed 700 μm. The laser cavity is a π-phase shifted waveguide Bragg grating sandwiched between two broadband waveguide Bragg gratings that enhance the pump radiation resonantly. For an initial simple and practical proof-of-concept demonstration, we considered a Raman waveguide laser that is pumped by an Er:ZBLAN MIR fiber-laser (λ=2.7 μm) and generates a wavelength of 3.15 mm. Our calculations show that for realistic value of propagation loss (1.0 dB/cm), a threshold pump power of less than 10 mW is readily achievable (for the design shown in Fig. 4(a)). We have also estimated an upper limit of 50 MHz for a pps-WBG silicon Raman laser fabricated based on a SOI waveguide with a loss of 1 dB /cm. The proposed design may pave the road for a new class of narrow-linewidth integrated silicon based MIR sources with potentially high impact on the development of low-cost MIR silicon photonics sensors.

**Funding:** National Science Foundation (NSF) Grant #1232263.

**Acknowledgment:** We thank Dr. Haisheng Rong of Intel's Photonics Technology Lab for helpful discussions on several aspects of this research.